\documentclass{article}

\usepackage[nonatbib, preprint]{neurips_2025}

\usepackage[utf8]{inputenc} 
\usepackage[T1]{fontenc}    
\usepackage{hyperref}       
\usepackage{url}            
\usepackage{booktabs}       
\usepackage{amsfonts}       
\usepackage{nicefrac}       
\usepackage{microtype}      
\usepackage{xcolor}         
\usepackage{xspace}
\usepackage{graphicx}
\usepackage{enumitem}
\usepackage{amsmath}
\usepackage{amssymb}
\usepackage{syntax}
\usepackage{mdframed}
\usepackage{svg}
\usepackage{subcaption}

\usepackage{hyperref}
\usepackage{pdflscape}

\usepackage[natbib=true, style=numeric,sorting=none]{biblatex}
\newcommand{\code}[1]{{\fontfamily{cmtt}\fontseries{m}\fontshape{n}\selectfont\small{#1}}}

\newcommand{\dsl}{{ai.txt}\xspace}

\newcommand{\robo}{{robots.txt}\xspace}

\definecolor{softred}{RGB}{180, 30, 30}
\newcommand{\kw}[1]{\textcolor{softred}{\texttt{`#1'}}}

\definecolor{darkgreen}{RGB}{0, 100, 0}
\newcommand{\ckw}[1]{\textcolor{darkgreen}{\texttt{#1}}}

\newmdenv[
  backgroundcolor=gray!20,
  linewidth=0pt,
  innerleftmargin=10pt,
  innerrightmargin=10pt,
  innertopmargin=10pt,
  innerbottommargin=10pt
]{coloredgrammar}

\title{\dsl{}: A Domain-Specific Language for Guiding AI Interactions with the Internet}

\author{%
Yuekang Li$^{1}$ \quad Wei Song$^{1}$ \quad Bangshuo Zhu$^1$ \quad Dong Gong$^1$ \quad Yi Liu$^2$ \quad \textbf{Gelei Deng}$^2$ \\ 
\quad \textbf{Chunyang Chen}$^3$\quad  \textbf{Lei Ma}$^{4,5}$ 
\quad  \textbf{Jun Sun}$^6$ \quad 
\textbf{Toby Walsh}$^1$ 
\quad \textbf{Jingling Xue}$^1$\\
$^1$University of New South Wales \quad $^2$Nanyang Technological University \\ 
$^3$Technical University of Munich\quad
$^4$University of Tokyo \quad $^5$University of Alberta\\
\quad $^6$Singapore Management University \\
\texttt{\{yuekang.li,wei.song1,bang.zhu,dong.gong,t.walsh,j.xue\}@unsw.edu.au}\\
\texttt{\{yi009,gdeng003\}@e.ntu.edu.sg} \quad
\texttt{chun-yang.chen@tum.de} \\
\texttt{ma.lei@acm.org} \quad
\texttt{junsun@smu.edu.sg}
}

\begin{document}

\maketitle

\begin{abstract}

We introduce \dsl{}, a novel domain-specific language (DSL) designed to explicitly regulate interactions between AI models, agents, and web content, addressing critical limitations of the widely adopted \robo{} standard. 
As AI increasingly engages with online materials for tasks such as training, summarization, and content modification, existing regulatory methods lack the necessary granularity and semantic expressiveness to ensure ethical and legal compliance. 
\dsl{} extends traditional URL-based access controls by enabling precise element-level regulations and incorporating natural language instructions interpretable by AI systems. 
To facilitate practical deployment, we provide an integrated development environment with code autocompletion and automatic XML generation. 
Furthermore, we propose two compliance mechanisms: XML-based programmatic enforcement and natural language prompt integration, and demonstrate their effectiveness through preliminary experiments and case studies. 
Our approach aims to aid the governance of AI-Internet interactions, promoting responsible AI use in digital ecosystems.

\end{abstract}

\section{Introduction}

With the emergence of large language models (LLMs) and their variants, such as large vision-language models (LVLMs), artificial intelligence (AI) has become increasingly integral to everyday life. These models have demonstrated remarkable capabilities across numerous fundamental tasks in natural language processing, image processing, and related domains. To leverage these advanced AI models, researchers and developers build AI agents—autonomous software systems designed to perform tasks independently or semi-autonomously on behalf of humans~\cite{Xi2023TheRA}. AI agents significantly amplify the effectiveness of underlying models by incorporating sophisticated programmatic logic, enabling them to automate intricate activities such as software development and maintenance~\cite{Yang2024SWEagentAI}.

As AI models and agents proliferate, their interactions with external environments, particularly the Internet, have expanded significantly. Among various interaction modalities, the engagement of AI with online resources represents a critical area for consideration. On the one hand, Internet-based content serves as valuable training and distillation data, enhancing the performance and adaptability of AI models. However, trained AI models and their associated agents actively interact with online content by utilizing, modifying, or generating new digital materials, thereby influencing the Internet ecosystem profoundly.

These interactions are currently underregulated, raising various practical and ethical concerns. For example, recent litigation initiated by The New York Times against OpenAI and Microsoft underscores legal challenges related to unauthorized use of copyrighted-protected content for training GPT-series models~\cite{nyt-openai}. Moreover, the use of distinctive stylistic references, such as ``Ghibli Style'' in GPT-4o image-generation, has raised additional concerns about copyright infringement~\cite{ghibli}. Ethical issues also emerge in the domain of AI-generated text summarization, including the potential erosion of deep comprehension, bias in summarization outcomes, and accountability complexities~\cite{ai-summary}. Consequently, establishing robust regulatory frameworks for governing AI-Internet interactions is essential.

\begin{figure}[t]
  \centering
  \includegraphics[width=1\linewidth]{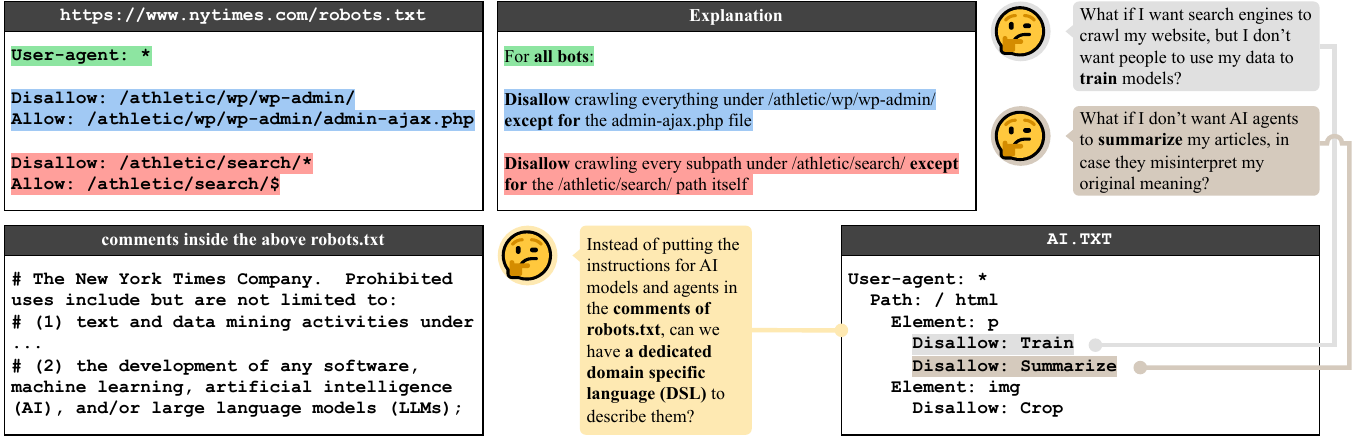}
  \caption{An example of \robo{} vs. \dsl{}.}
  \label{fig:example}
\end{figure}

Regulatory measures currently employed, such as the widely adopted \textit{\robo{}} standard~\cite{robots-google}, prove insufficient for addressing the complexities introduced by AI models and agents. 
Initially proposed by Martijn Koster in 1994~\cite{robots-orig}, the \robo{} standard enables website owners to guide web crawlers to access or exclude specific URLs through a simple textual configuration file typically located at \url{/robots.txt}, as exemplified by \url{https://www.nytimes.com/robots.txt}. 
Although \robo{} effectively instructs traditional web crawlers on URL-based access permissions, it lacks the granularity and semantic expressiveness necessary to regulate nuanced behaviors of contemporary AI systems. 
For example, Figure~\ref{fig:example} illustrates the \robo{} file of The New York Times, effectively managing crawler access at the URL level. 
Yet, despite clearly expressed concerns about the use of their articles for training LLMs, such restrictions cannot be formally enforced using the current \robo{} syntax and are instead confined to informal textual comments.
Likewise, issues surrounding the accuracy of article summarization and the fidelity of interpretation remain unresolvable through existing technical means.

Consequently, the existing \robo{} standard does not adequately manage and constrain the diverse and sophisticated interactions between AI agents and the web content. 
Addressing these limitations necessitates developing an enhanced regulatory framework, explicitly designed to encode complex semantic constraints and ethical considerations, ensuring more precise and meaningful governance of AI interactions on the Internet.

To address this gap, we propose a novel domain-specific language (DSL) named \dsl{}. 
The design of \dsl{} adheres to the core principles of simplicity, clarity, consistency, and functionality. 
Figure~\ref{fig:example} provides an illustrative example of an \dsl{} file. 
The syntax of \dsl{} closely resembles that of \robo{}, ensuring human readability while extending its capabilities to explicitly regulate various actions performed by AI agents.
Importantly, \dsl{} facilitates more fine-grained control over regulated online content compared to conventional methods. Instead of managing AI actions based solely on website paths, \dsl{} enables specific regulation of individual elements within online content. 
For example, referring to the file shown in Figure~\ref{fig:example}, actions such as \code{Train} and \code{Summarize} can be explicitly disallowed for HTML paragraphs, whereas the \code{Crop} action can be explicitly prohibited for images. 
Additionally, beyond simple permission management, \dsl{} supports the provision of natural language instructions intended for AI agents capable of interpreting such guidance.

To make it easy to develop and maintian \dsl{} files, we implemented an integrated development environment (IDE) for \dsl{} using JetBrains' Meta Programming System (MPS)~\cite{mps}. 
The IDE simplifies the development of \dsl{} files by providing code hints and autocompletion. Furthermore, an XML generation feature is integrated into the IDE, enabling automatic conversion of \dsl{} files into XML format for straightforward parsing using existing mature parser libraries.

To ensure AI agents adhere to \dsl{} specifications, we propose two complementary compliance mechanisms. 
First, AI agents can directly parse the generated XML files to enforce regulations programmatically. 
Second, AI agents equipped with natural language processing (NLP) capabilities can interpret the plain text of \dsl{} files as actionable instructions, incorporating them into their prompt-based regulation mechanisms. 
Preliminary experimental results indicate that both methods effectively enable AI agents to follow \dsl{} directives. 
Additionally, case studies confirm the expressiveness of \dsl{} to comprehensively cover diverse regulatory scenarios.

In summary, in this work we make the following key contributions:

\begin{itemize}[leftmargin=*]
\item We propose \dsl{}, a novel DSL designed to explicitly instruct and regulate interactions between AI agents and the Internet, potentially exerting significant influence over the broader digital ecosystem.
\item We develop an IDE tailored for creating \dsl{} files, with autocompletion support and integrated XML generation capabilities.
\item We present two complementary approaches which enable AI agents to comply with \dsl{} regulations, each accompanied by an implementation framework designed to integrate seamlessly with existing AI agent systems.
\item Through preliminary experiments, we validate the flexibility and expressiveness of \dsl{} in accommodating various practical use cases.
\end{itemize}

This work is accompanied by a project website that provides additional explanations and the source code for the \dsl{} tools: \url{https://sites.google.com/view/ai-txt/home}.

\section{Background and Related Work}

\subsection{\robo{}}



\robo{}~\cite{rfc-9309}, formally defined under the Robots Exclusion Protocol, is a widely adopted standard which allows webmasters to specify which sections of their websites are accessible to web crawlers and search engines. 
Initially introduced by Martijn Koster in 1994~\cite{robots-orig}, it manifests as a simple text file placed at the root directory of the website. 
The file includes directives such as \texttt{User-agent}—identifying specific crawlers—and \texttt{Disallow}—indicating URLs excluded from indexing. 
Despite its simplicity, the original \robo{} specification suffered from inherent ambiguity, resulting in inconsistent interpretations and implementations across different web crawlers. 
To mitigate these discrepancies, Google initiated a Request for Comment (RFC) in 2019, culminating in the publication of RFC-9309 in 2022~\cite{rfc-9309}.

Whilst \robo{} serves as a regulatory guideline for web crawlers, it does not inherently enforce compliance. 
Effective regulation under \robo{} necessitates supplementary anti-crawler techniques to enforce adherence~\cite{Park2006SecuringWS,Jacob2012PUBCRAWLPU}. 
Similarly, the proposed \dsl{} is designed to allow website administrators to declare intended regulations for AI agents explicitly, rather than to enforce them directly. 
We outline strategies for building AI agents that adhere to \dsl{} guidelines, while leaving the enforcement of these directives through \textit{anti-AI} techniques as a direction for future research.
In contrast to \robo{}, which primarily targets web crawlers, \dsl{} provides the granularity and semantic precision necessary to instruct the complex behaviors of contemporary AI systems effectively. 
Thus, inspired by the regulatory approach of \robo{}, \dsl{} establishes a structured framework to precisely govern interactions between AI agents and the Internet.

\subsection{AI Regulation}
In recent years, numerous countries have issued documents to regulate AI~\cite{EU-AI-Act, White-House-Fact-Sheet, Japan-AI-Regulation-Report, CA-AI-Regulation, CA-AI-Regulation-Review, CN-AI-Regulation-Report, LA-AI-Regulation}. 
The \emph{EU AI Act}~\cite{EU-AI-Act}, proposed by the European Commission in 2021, is widely recognized as the first comprehensive legislation which specifically addresses AI governance. 
The act emphasizes ensuring that AI systems deployed within the European Union are safe, transparent, traceable, non-discriminatory, and environmentally sustainable. 
It categorizes AI systems based on risk levels, establishing regulations accordingly. 
Following the \emph{EU AI Act}, many countries issued analogous regulatory frameworks in 2023~\cite{White-House-Fact-Sheet, Japan-AI-Regulation-Report, CA-AI-Regulation, CA-AI-Regulation-Review, CN-AI-Regulation-Report, LA-AI-Regulation}, which also aim to ensure the safe, secure, and trustworthy development, deployment, and utilization of AI. 
These regulatory documents mark a critical milestone in addressing both the potential risks and benefits associated with AI technologies.

Collectively, these policies set comprehensive standards intended to safeguard various aspects of AI applications, notably powerful LLM-based systems. 
They address concerns such as data privacy, information governance, social equity, and environmental impacts across critical domains including education, healthcare, public privacy, civil rights, and equity. 
For example, the White House issued the \emph{Executive Order on Safe, Secure, and Trustworthy Artificial Intelligence}~\cite{White-House-Fact-Sheet}, establishing a rigorous legal framework in the United States aimed at promoting safety, security, and ethical considerations within AI applications.

However, despite their comprehensive nature, these governmental regulatory guidelines~\cite{EU-AI-Act, White-House-Fact-Sheet, Japan-AI-Regulation-Report, CA-AI-Regulation, CA-AI-Regulation-Review, CN-AI-Regulation-Report, LA-AI-Regulation} provide only broad, overarching principles rather than precise technical specifications or detailed compliance monitoring instructions~\cite{Global-Policies, EU-AI-Act, US-AI-Regulation}. 
Consequently, there remains a critical need for technical frameworks and methodologies capable of explicitly guiding and verifying compliance with these emerging regulations.


\section{Design of \dsl{}}



\subsection{Design Principles} \label{sec:design-principle}

\begin{figure}[!h]
  \centering
  \includegraphics[width=1\linewidth]{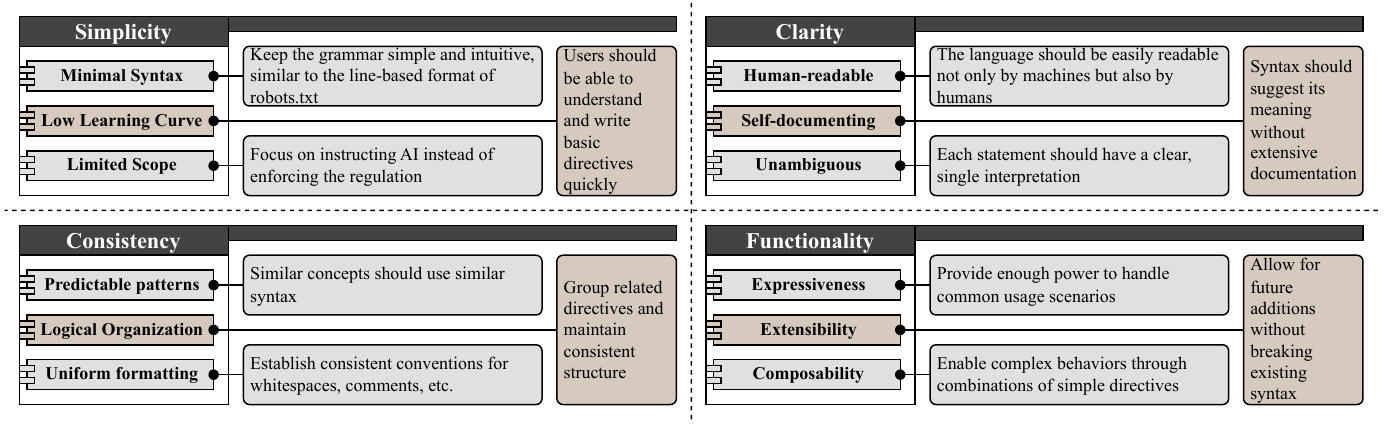}
  \caption{The design principles for \dsl{}.}
  \label{fig:design-principle}
\end{figure}

When designing \dsl{}, we focus on four core principles that ensure effectiveness and adoption. 
\textbf{Simplicity} is paramount; we aim to create a minimal syntax with a low learning curve that addresses a limited problem scope. 
\textbf{Clarity} requires human-readable, self-documenting code where directives have unambiguous meanings. 
\textbf{Consistency} builds user confidence through predictable patterns, logical organization, and uniform formatting conventions. 
Finally, \textbf{functionality} balances expressiveness to handle common use cases with extensibility for future growth and composability to create complex behaviors from simple elements. 
Together, these principles create a DSL that users can quickly understand and implement while providing enough power to effectively solve the domain-specific challenges of AI regulation.

\subsection{Language Design}

We present the syntax design of \dsl{} using the Extended Backus–Naur Form (EBNF), adhering to the style recommended by the W3C~\cite{w3cebnf}. 
Additionally, we provide corresponding railroad diagrams to visually illustrate the structures of the fundamental components of the language. 
In this section, we introduce the key syntactic concepts, while comprehensive details are provided in Appendix~\ref{app:syntax}.

\begin{figure}[!h]
  \centering
  \includegraphics[width=1\linewidth]{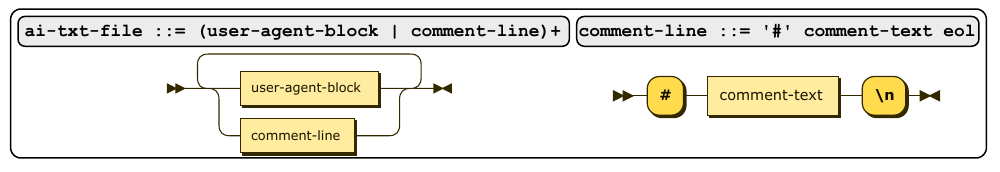}
  \caption{The \dsl{} file and comment in EBNF and railroad diagram.}
  \label{fig:bnf-file}
\end{figure}

At the topmost level, an \dsl{} file consists of one or more \code{user-agent} blocks and optional comment lines, as illustrated in Figure~\ref{fig:bnf-file}. 
Although empty files are syntactically permissible, they serve no practical purpose. 
Similarly to conventions in \robo{} and languages such as Python, comment lines in \dsl{} begin with the character \code{\#}, followed by arbitrary explanatory text.

\begin{figure}[!h]
  \centering
  \includegraphics[width=1\linewidth]{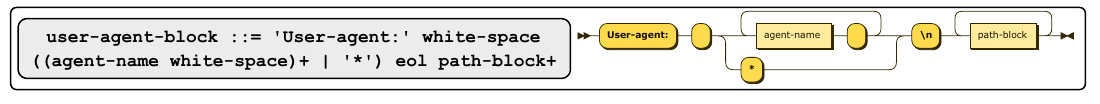}
  \caption{The user-agent block in EBNF and railroad diagram.}
  \label{fig:bnf-useragent}
\end{figure}

The \code{user-agent} blocks specify the regulatory instructions applicable to one or more AI models or agents. 
Figure~\ref{fig:bnf-useragent} illustrates the detailed syntax structure. 
Each \code{user-agent} block begins with an information line, followed by one or more \code{path} blocks. 
The information line starts with the keyword \code{User-agent:}, followed by the names of the agents to be regulated. 
This list of agent names includes either multiple identifiers separated by whitespace or the wildcard character \code{*}, indicating applicability to all agents. 
Valid agent names consist of alphanumeric characters (\code{a-zA-Z0-9}) and underscores (\code{_}).

\begin{figure}[!h]
  \centering
  \includegraphics[width=1\linewidth]{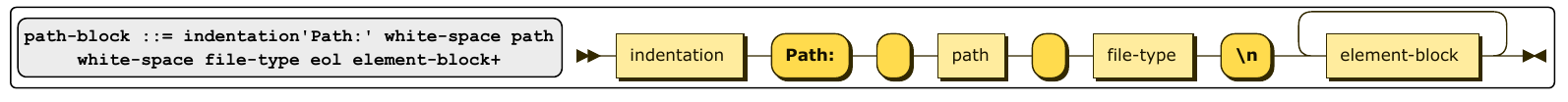}
  \caption{The path block in EBNF and railroad diagram.}
  \label{fig:bnf-path}
\end{figure}

The \code{path} blocks specify the details of individual paths subject to regulation, each beginning with a forward slash character (\code{/}). 
Figure~\ref{fig:bnf-path} presents the detailed syntax. 
Each \code{path} block is indented exactly once and starts with an information line, followed by one or more subsequent \code{element} blocks. 
The information line begins with the keyword \code{Path:}, followed by the specific path value and its corresponding \code{file-type}. The specified paths are relative to the website root. 
For example, given the URL \code{https://en.wikipedia.org/wiki/Robots.txt}, the actual path is represented as \code{/wiki/Robots.txt}. 
The \code{file-type} denotes the format of the file associated with the specified path. 
Currently supported \code{file-type} values include \code{html}, \code{json}, and \code{xml}.

\begin{figure}[!h]
  \centering
  \includegraphics[width=1\linewidth]{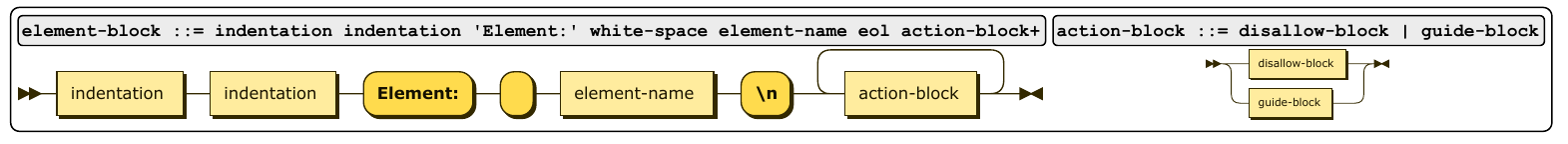}
  \caption{The element block and action block in EBNF and railroad diagram.}
  \label{fig:bnf-element}
\end{figure}

The \code{element} blocks define the specific elements within a given path that are subject to regulation. 
The detailed syntax is illustrated in Figure~\ref{fig:bnf-element}. 
Each \code{element} block is indented twice and begins with an information line, followed by one or more \code{action} blocks.
The information line starts with the keyword \code{Element:}, followed by the \code{element-name}. 
With the exception of the wildcard character \code{*}, the format of \code{element-name} must be consistent with the \code{file-type} specified in the parent \code{path} block. 
If the \code{file-type} is \code{json} or \code{xml}, the \code{element-name} should correspond to the subobject name, using dot notation where appropriate. 
If the \code{file-type} is \code{html}, the \code{element-name} should conform to the syntax of CSS selectors~\cite{cssselector}, enabling precise identification of DOM elements.

Each \code{action} block within an \code{element} block can either be a \code{Disallow} directive or a \code{Guide} directive, as shown in Figure~\ref{fig:bnf-element}.

\begin{figure}[!h]
  \centering
  \includegraphics[width=1\linewidth]{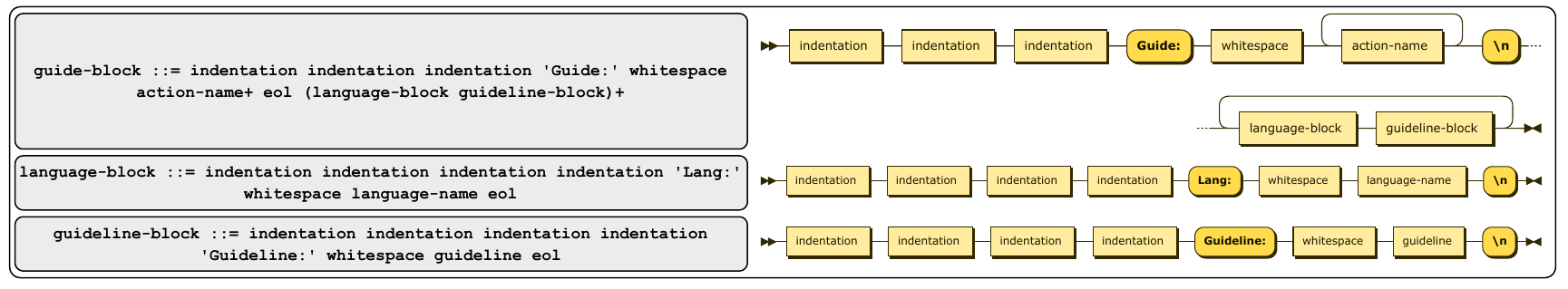}
  \caption{The guide block in EBNF and railroad diagram.}
  \label{fig:bnf-guide}
\end{figure}

\begin{figure}[!h]
  \centering
  \includegraphics[width=1\linewidth]{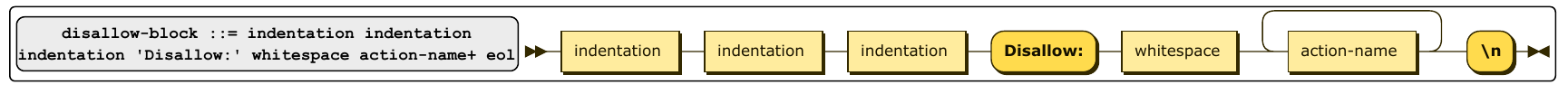}
  \caption{The disallow block in EBNF and railroad diagram.}
  \label{fig:bnf-disallow}
\end{figure}

The \code{guide} blocks specify actions that are permitted but require additional instructions for AI agents when applied to the corresponding path and element. 
The detailed syntax is shown in Figure~\ref{fig:bnf-disallow}. 
Each \code{guide} block is indented three times and consists of an information line, followed by one or more pairs of \code{language-block} and \code{guideline-block}.
The information line begins with the keyword \code{Guide:}, followed by a list of actions. 
This list can consist of predefined action names separated by whitespace or the wildcard character \code{*}, which denotes all supported actions (see Section~\ref{sec:actions}). 
For each specified action, website owners may provide language-specific textual guidelines intended to inform agent behavior.
The language identifiers should conform to ISO 639 language codes~\cite{iso-639}, such as \code{en-US}, \code{en-UK}, or \code{en-AU}. 
Each corresponding \code{guideline} is a plaintext instruction written in the specified language. 
These guidelines can be integrated into the agent's prompt to guide its behavior for the associated action. 
For example, for the \code{Summarize} action, a guideline in \code{en-US} might be: ``Please keep the first line of each paragraph in the summarization.''

The \code{disallow} blocks specify the actions that are prohibited for AI agents on the corresponding path and element. 
The detailed syntax is illustrated in Figure~\ref{fig:bnf-disallow}. 
Each \code{disallow} block is indented three times and consists of a single line. 
This line begins with the keyword \code{Disallow:}, followed by a list of actions, formatted similarly to the action list in the \code{guide} block.

\subsection{Regulated Actions} \label{sec:actions}

\begin{figure}[!h]
  \centering
    \begin{subfigure}[t]{0.5\textwidth}
        \centering
        \includegraphics[width=\linewidth]{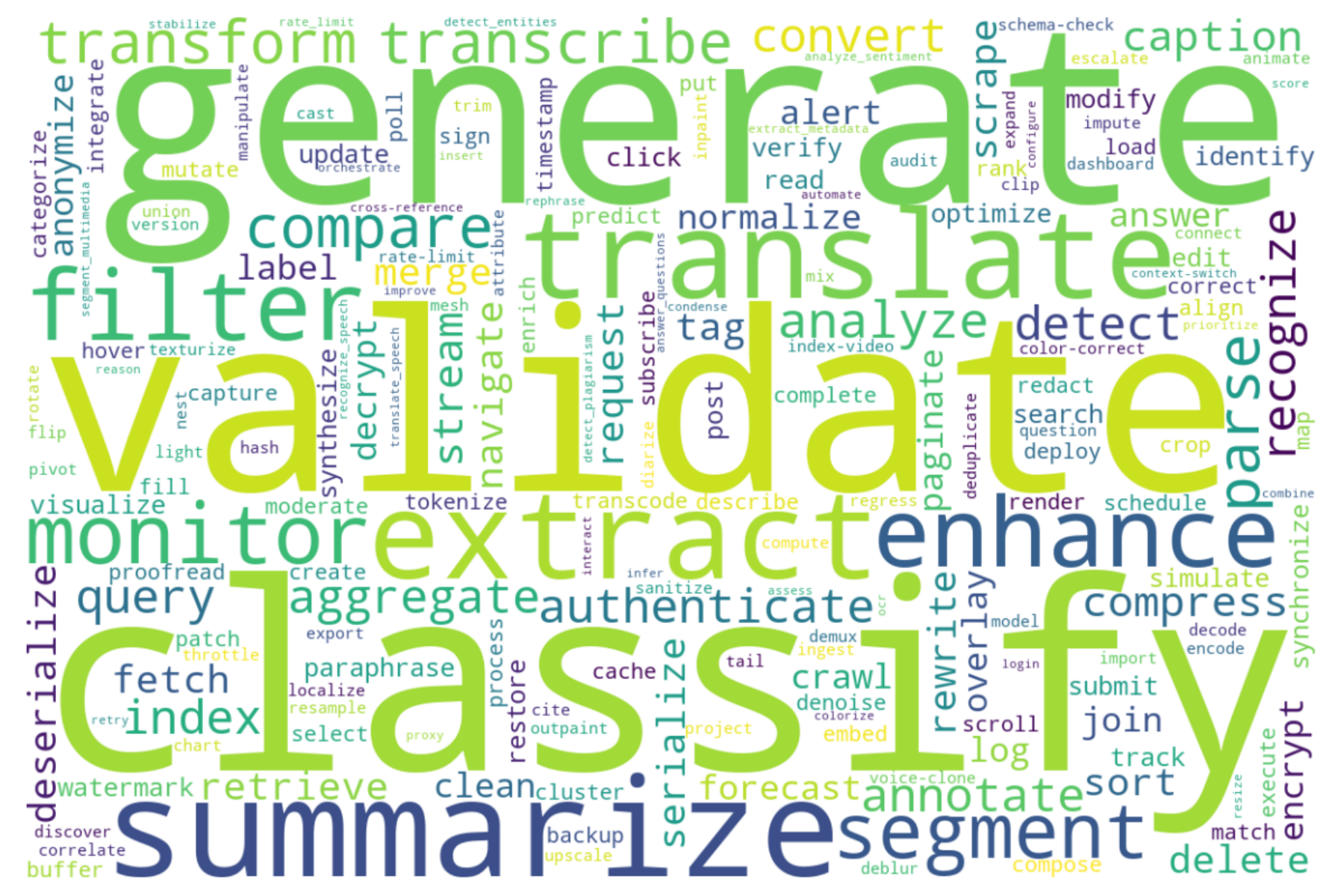}
        \caption{Word cloud from the answers of LLMs}
    \end{subfigure}%
    ~ 
    \begin{subfigure}[t]{0.5\textwidth}
        \centering
        \includegraphics[width=\linewidth]{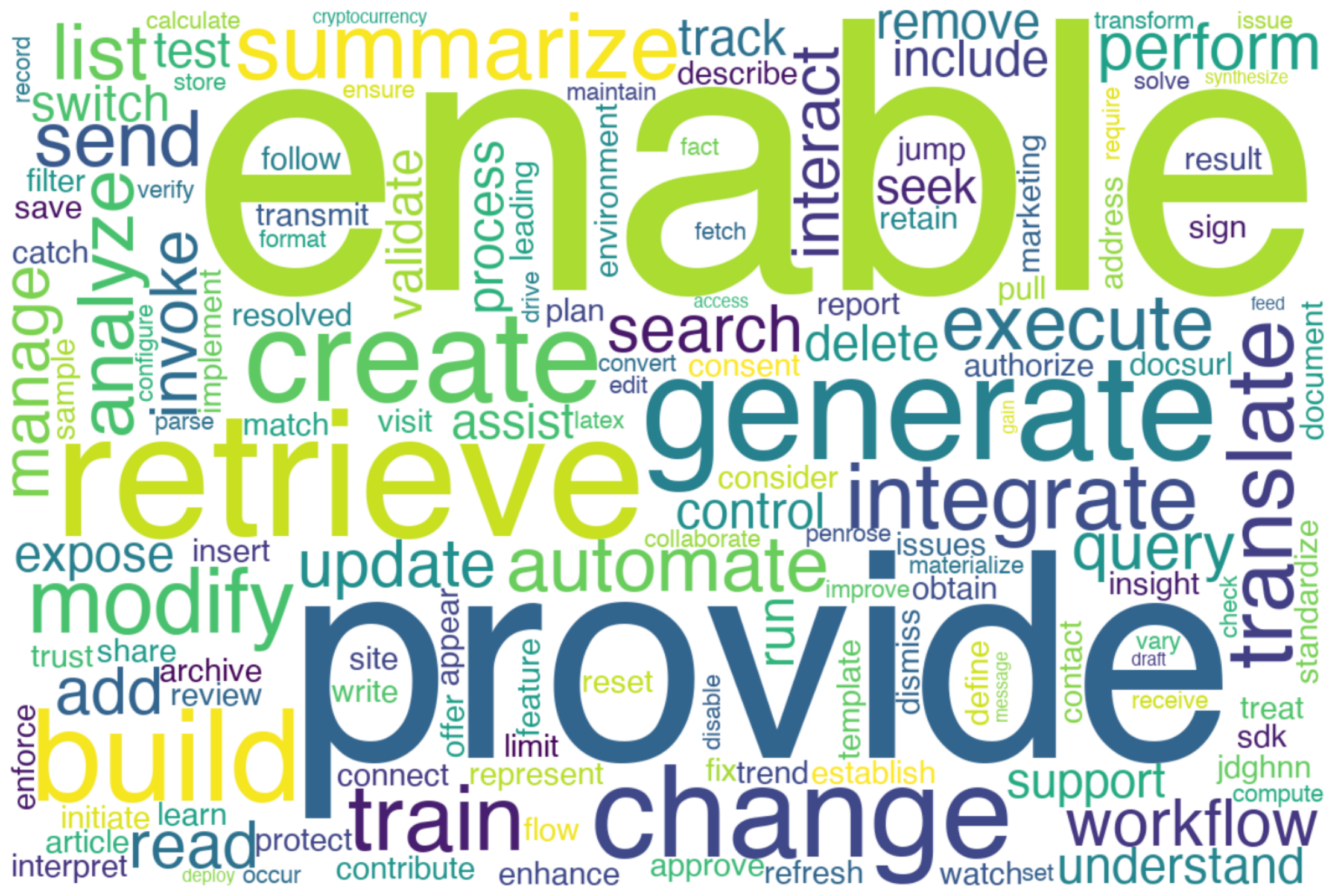}
        \caption{Word cloud from the MCP server descriptions}
    \end{subfigure}
    \caption{Collected word clouds}
  \label{fig:wordcloud}
\end{figure}

We aim to systematically compile a comprehensive set of actions that AI agents can perform on various types of web content, including HTML, JSON, and XML, covering textual, image, and multimedia data. 
To achieve this, we initially constructed a preliminary set of actions using two complementary methods: (1) querying multiple large language models (LLMs), such as GPT series of models, to obtain enumerations of actionable verbs relevant to AI interactions with web content; and (2) crawling textual descriptions from 3,592 MCP server webpages to extract action-related terms.
Figure~\ref{fig:wordcloud} shows the world clouds of the action verbs we initially collected.
Subsequently, we manually filtered out overly broad or ambiguous terms that lacked sufficient specificity. 
Additionally, we reviewed existing regulatory documents and AI governance frameworks~\cite{EU-AI-Act, White-House-Fact-Sheet, Japan-AI-Regulation-Report, CA-AI-Regulation, CA-AI-Regulation-Review, CN-AI-Regulation-Report, LA-AI-Regulation} to gain insight into standardized terminologies and best practices in defining actionable capabilities. 
Through iterative refinement and cross-validation of these sources, we finalized a curated, precise list of actionable verbs suitable for rigorous research and practical implementation in AI agent regulation.

The finalized list of actions is detailed as follows:

\begin{itemize}[leftmargin=*]
    \item \textbf{Analyze}: Perform complex reasoning, derive insights, generate hypotheses, or draw conclusions based on the content, going beyond surface description or summarization. Content marked with \code{Disallow: Analyze} must not be used as a basis for complex analysis or insight generation by the AI agent.
    \item \textbf{Cite}: Attribute or acknowledge the original source of content elements (text, image, audio, video). Elements marked with \code{Disallow: Cite} must not have the original source explicitly attributed in AI-generated responses.
    \item \textbf{Clip}: Extract shorter segments or subsets of multimedia content (audio and video). Content marked with \code{Disallow: Clip} must not be shortened; the entire original length must be preserved when returned by the AI agent.
    \item \textbf{Describe}: Generate objective and surface-level explanations or interpretations of content elements (text, image, audio, video), such as identifying visual objects or explaining literal meanings. Content marked with \code{Disallow: Describe} must not be described by the AI agent.
    \item \textbf{Evaluate}: Perform assessments or judgments regarding the quality, sentiment, bias, toxicity, or other evaluative metrics of content elements. Content marked with \code{Disallow: Evaluate} must not undergo evaluative analysis.
    \item \textbf{Extract}: Automatically retrieve structured or semi-structured information from web content (text, JSON, XML, HTML), supporting tasks such as data mining, entity recognition, or metadata extraction. Content marked with \code{Disallow: Extract} must not have information extracted.
    \item \textbf{Index}: Process and store content or its representations (like keywords or vector embeddings) in a searchable index used by the AI agent for retrieval or similarity matching. Content marked with \code{Disallow: Index} must not be included in AI-accessible search indexes.
    \item \textbf{Manipulate}: Modify or edit multimedia content (images, audio, videos), including creating deepfakes, impersonating individuals, altering stylistic elements (e.g., filters, style transfer), cropping subjects, remixing, and performing digital edits. Content marked with \code{Disallow: Manipulate} must remain entirely unaltered if used by the AI agent.
    \item \textbf{Rephrase}: Reformulate or restate textual content into alternative phrasing. Textual content marked with \code{Disallow: Rephrase} must not be rephrased and must instead be quoted verbatim if returned in AI-generated responses.
    \item \textbf{Return}: Directly return or incorporate original content elements (text, image, audio, video) into AI-generated outputs. Content marked with \code{Disallow: Return} must be excluded entirely from AI-generated responses.
    \item \textbf{Summarize}: Produce concise summaries reflecting deeper meanings or underlying themes of content elements (text, image, audio, video), addressing interpretive or thematic questions rather than surface-level descriptions. Content marked with \code{Disallow: Summarize} must not be summarized.
    \item \textbf{Train}: Incorporate content elements (text, image, audio, video) into datasets used for AI model training. Content explicitly tagged with \code{Disallow: Train} must be excluded from all training datasets.
    \item \textbf{Transcribe}: Accurately convert spoken language from audio or video content into written textual form. Content marked with \code{Disallow: Transcribe} must not be transcribed.
    \item \textbf{Translate}: Accurately convert textual content from one natural language into another. Content marked with \code{Disallow: Translate} must not be translated.
\end{itemize}

These actions are explicitly defined based on their clarity, specificity, and practical applicability within regulatory frameworks governing AI agents interacting with web-based content. 
Notably, the action list is designed to be extensible and may be modified or expanded to incorporate new actions, in accordance with the design principles outlined in Section~\ref{sec:design-principle}.



\section{Using \dsl{} for Regulating AI}

\begin{figure}[!h]
  \centering
    \begin{subfigure}[t]{0.5\textwidth}
        \centering
        \includegraphics[width=\linewidth]{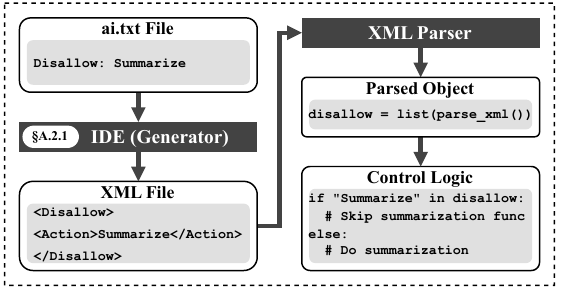}
        \caption{Programmatic enforcement}
    \end{subfigure}%
    ~ 
    \begin{subfigure}[t]{0.5\textwidth}
        \centering
        \includegraphics[width=\linewidth]{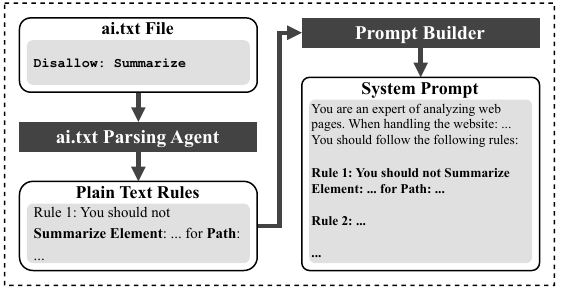}
        \caption{Prompt-level enforcement}
    \end{subfigure}
    \caption{Usage strategies of \dsl{}}
  \label{fig:usage}
\end{figure}

We propose two complementary approaches for enabling AI models or agents to comply with the behavioral constraints specified in \dsl{}. 
Figure~\ref{fig:usage} illustrates the two strategies.
The first approach is \emph{programmatic enforcement}. 
In this method, an \dsl{} file is compiled into a structured representation, such as XML or other standardized formats with well-supported parsers. 
Developers can then integrate rule-checking logic into the AI agent's execution pipeline. 
For instance, if the parsed file contains a directive such as \code{Disallow: Summarize}, and the agent is about to invoke a summarization function, the control flow should be programmatically altered to prevent its execution. 
This approach offers strict, deterministic enforcement and is suitable for scenarios where agent behavior must be tightly controlled at the system level.

The second approach is \emph{prompt-level enforcement}, which leverages the interpretive capabilities of LLMs. 
Here, the \dsl{} file is treated as plain text input and interpreted by an AI agent to extract a list of behavioral rules. 
These extracted rules are then embedded into the prompts of downstream, regulated AI agents, guiding their behavior in a flexible and context-aware manner. 
This approach is particularly useful for natural language-based agents operating in less structured environments, and it enables compliance without modifying the internal logic of the AI system. 
Together, these approaches provide complementary mechanisms for aligning AI agent behavior with declarative, human-readable policies.

\section{Evaluation}

We present a preliminary experimental design to evaluate the extent to which contemporary AI models and agents can adhere to user-defined behavioral constraints. 

Specifically, we distinguish between the two enforcement strategies: programmatic enforcement—where the execution logic of AI agents is explicitly governed by external rule parsers—and prompt-level enforcement, which relies on embedding regulatory instructions directly within the input prompts. 
Given that programmatic enforcement can deterministically prevent disallowed actions, our focus is on assessing the effectiveness of prompt-level enforcement, which is inherently more susceptible to misinterpretation or circumvention by the model. 
Initial experiments suggest that state-of-the-art models such as GPT-4o demonstrate high compliance under prompt-level guidance, with minimal observed violations. 
However, these results are preliminary, and further testing is required to systematically examine model behavior under adversarial prompts, ambiguous instructions, and more complex regulatory conditions.

\section{Conclusion}

In this paper, we introduce \dsl{}, a domain-specific language designed to regulate the behavior of AI models and agents when interacting with web content. The language design is grounded in four key principles—\textit{simplicity}, \textit{clarity}, \textit{consistency}, and \textit{functionality}—which collectively ensure that \dsl{} remains human-readable and machine-actionable. 
We rigorously define the syntax and semantics of \dsl{}, which enables unambiguous specification of permissible and prohibited actions across diverse content types. 
Furthermore, we propose and implement two complementary enforcement strategies: programmatic enforcement through structured rule parsing, and prompt-level enforcement using natural language descriptions of regulatory constraints. 
Our experimental evaluation assesses the extent to which state-of-the-art AI models adhere to \dsl{}-based instructions, revealing that while high-performing models like GPT-4o exhibit strong compliance in standard scenarios, challenges remain in complex or ambiguous contexts.

\printbibliography

\appendix

\section{Technical Appendices and Supplementary Material}

\subsection{Syntax Details for \dsl{}} \label{app:syntax}

Here are the complete details about the syntax of \dsl{} in Figure~\ref{fig:complete}.

\begin{figure}[t]
\footnotesize

\centering
\[
\begin{array}{rcl}
\textit{ai-txt-file} &::= & (\textit{user-agent-block} \ | \ \kw{\#} \ \textit{comment-text} \ \textit{eol})+ \\[1.5ex]

\textit{comment-line} &::= & \kw{\#} \ \textit{comment-text} \ \textit{eol} \\

\textit{user-agent-block} &::=& \kw{User-agent:} \ \textit{white-space} \ ((\textit{agent-name} \ \textit{white-space})+ \ | \ \kw{\*}) \ \textit{eol} \\ &|& \textit{path-block}+ \\
\textit{path-block} &::=& \textit{indentation} \ \kw{Path:} \ \textit{white-space} \ \textit{path} \ \textit{white-space} \ \textit{file-type} \ \textit{eol} \\ &|&  \textit{element-block}+ \\
\textit{element-block} &::=& \textit{indentation} \ \textit{indentation} \ \kw{Element:} \ \textit{white-space} \textit{element-name} \ \textit{eol} \\ &|& \textit{action-block}+ \\
\textit{action-block} &::=& \textit{disallow-block} \ | \ \textit{guide-block} \\
\textit{disallow-block} &::=& \textit{indentation} \ \textit{indentation} \ \textit{indentation} \ \kw{Disallow:} \ \textit{whitespace} \ \textit{action-name}+ \ \textit{eol} \\
\textit{guide-block} &::=& \textit{indentation} \ \textit{indentation} \ \textit{indentation} \ \kw{Guide:} \ \textit{whitespace} \ \textit{action-name}+ \ \textit{eol} \\ &|& (\textit{language-block} \ \textit{guideline-block})+ \\
\textit{language-block} &::=& \textit{indentation} \ \textit{indentation} \ \textit{indentation} \ \textit{indentation} \kw{Lang:} \ \textit{whitespace} \ \textit{language-name} \ \textit{eol} \\
\textit{guideline-block} &::=& \textit{indentation} \ \textit{indentation} \ \textit{indentation} \ \textit{indentation} \ \kw{Guideline:} \ \textit{whitespace} \ \textit{guideline} \ \textit{eol} \\

\textit{file-type} &::=& \kw{html} | \kw{json} | \kw{xml} \\
\textit{eol} &::=& \kw{\textbackslash n} \\
\textit{white-space} &::=& \kw{ } \\
\textit{indentation} &::=& (\textit{white-space} \ \textit{white-space} \ ( \textit{white-space} \ \textit{white-space} )?) \ | \ \kw{\textbackslash t} \\

\textit{agent-name} &::=& \ckw{[a-zA-Z0-9_]+} \\

\textit{path} &::=& \kw{/} \ (\textit{path-segment} \ \kw{/})* (\textit{path-segment})? \\
\textit{path-segment} &::=& \ckw{[a-zA-Z0-9_.\&\%/~:@-]+} \\

\end{array}
\]
\caption{Complete EBNF grammar for \dsl{}.}
\label{fig:complete}
\end{figure}
    
    


  


\subsection{Implementation Details}

\subsubsection{IDE for \dsl{}}

We developed an integrated development environment (IDE) for \dsl{} using the Meta Programming System (MPS) by JetBrains~\cite{mps}. 
The IDE features a code editor with autocompletion and various constraint checkers—for example, validating whether a given \code{path} adheres to the required format. 
Additionally, the IDE includes a generator that converts \dsl{} files into XML format to facilitate parsing with existing tools. 
A demonstration video showcasing the IDE’s functionality is available on our project website~\cite{our-site}.

\subsubsection{Agent Framework for \dsl{}}

The source code of the AI Agents and our suggested framework of using \dsl{} in development will be released on our website~\cite{our-site}.

\end{document}